\documentclass[twocolumn,preprintnumbers,amsmath,amssymb]{revtex4}

\usepackage{graphicx}% Include figure files
\usepackage{dcolumn}% Align table columns on decimal point
\usepackage{bm}% bold math
\usepackage{setspace}

\begin{document}

\title{Long electron dephasing length and disorder-induced spin-orbit coupling
in indium tin oxide nanowires}

\author{Yao-Wen Hsu,$^1$ Shao-Pin Chiu,$^1$ An-Shao Lien,$^1$ and
Juhn-Jong Lin$^{1,2,}$}

\email{jjlin@mail.nctu.edu.tw}

\address{$^1$Institute of Physics, National Chiao Tung University,
Hsinchu 30010, Taiwan}

\address{$^2$Department of Electrophysics, National Chiao Tung
University, Hsinchu 30010, Taiwan}

\date{\today}

\begin{abstract}

We have measured the quantum-interference magnetoresistances in two single
indium tin oxide (ITO) nanowires between 0.25 and 40 K, by using the
four-probe configuration method. The magnetoresistances are compared with the
one-dimensional weak-(anti)localization theory to extract the electron
dephasing length $L_\varphi$. We found, in a 60-nm diameter nanowire with a
low resistivity of $\rho$(10 K) = 185 $\mu \Omega$ cm, that $L_\varphi$ is
long, increasing from 150 nm at 40 K to 520 nm at 0.25 K. Therefore, the
nanowire reveals strict one-dimensional weak-localization effect up to several
tens of degrees of Kelvin. In a second 72-nm diameter nanowire with a high
resistivity of $\rho$(10 K) = 1030 $\mu \Omega$ cm, the dephasing length is
suppressed to $L_\varphi$(0.26 K) = 200 nm, and thus a crossover of the
effective device dimensionality from one to three occurs at about 12 K. In
particular, disorder-induced spin-orbit coupling is evident in the latter
sample, manifesting weak-antilocalization effect at temperatures below $\sim$
4 K. These observations demonstrate that versatile quantum-interference
effects can be realized in ITO nanowires by controlling differing levels of
atomic defects and impurities.

\end{abstract}



\maketitle

\section{Introduction}

In weakly disordered metals and semiconductors (all real conductors are
disordered), quantum-interference electron transport plays vital roles at
liquid-helium temperatures \cite{Imry}. The quantum-interference phenomena,
which originate from the inherent wave nature of the conduction electrons,
become progressively pronounced as the temperature is lowered and the
effective device dimensionality is reduced. In this regard, metal and doped
semiconductor nanowires could provide a rich platform for probing novel,
nonclassical properties and functionalities. In a random potential, the
conduction electrons undergo Brownian-type diffusive motion, leading to
various quantum-interference phenomena, including the weak-(anti)localization
effect, Aharonov-Bohm oscillations, universal conductance fluctuations, and
persistent currents \cite{Nazarov}. Previously, such quantum-interference
behaviors have mostly been investigated by utilizing submicron devices
fabricated via the `top-down' electron-beam lithography technique
\cite{Washburn86}.

Compared with the temperature dependence of the resistance rise due to
weak-localization and electron-electron interaction effects at low
temperatures \cite{Altshuler87}, magnetoresistances induced by small
externally applied magnetic fields can provide very quantitative information
about the inelastic electron (electro-phonon and electron-electron) scattering
and spin-flip (spin-orbit and magnetic spin-spin) scattering mechanisms in a
given nanowire device \cite{Lin-jpcm02}. Therefore, systematic and careful
measurements of magnetoresistances in the weak-(anti)localization effect are
indispensable and highly desirable for learning the responsible microscopic
electron relaxation processes \cite{dephasing}. Reliable knowledge about the
electron relaxation processes can help in paving the way for the future
realization of sensitive and versatile quantum-interference nanoelectronics
and spintronics. Since as-grown metal and semiconductor nanowires often
contain high levels of point defects \cite{Chiu-nano09ITO,Chiu-nano09ZnO}, it
is natural to expect that quantum-interference effects should be pronounced in
many `bottom-up' nanostructures.

Among the numerous metallic nanowires, Sn-doped In$_2$O$_3$ (indium tin oxide
or ITO) nanowires have recently been studied \cite{Wan06,LinD07,Dwyer09}. The
interest in these nanoscale structures arises from the fact that the parent
ITO material is a transparent conducting oxide. Moreover, the charge carriers
in this material are established to be essentially free-electron-like
\cite{Mryasov01}. That is, the charge carriers behave like an ideal free
electron gas with an effective mass of $m^\ast \approx 0.4\,m$ and a small
Fermi energy $E_F \lesssim$ 1 eV, where $m$ is the free electron mass. The
thermoelectric power reveals a linear diffusive term from 300 K all the way
down to liquid-helium temperatures, while the temperature dependence of
resistivity is well described by the Bloch-Gr\"uneisen law at not too low
temperatures \cite{Li-JAP04}. The carrier concentration in highly conductive
ITO materials falls in the range $10^{20}$--$10^{21}$ cm$^{-3}$
\cite{Tahar98}, and the room temperature resistivity can be as low as $\sim$
100--200 $\mu \Omega$ cm \cite{Wan06}. Therefore, ITO nanowires appear to be
attractive candidates for in-depth investigations of the quantum-interference
electron transport. Surprisingly, to the best of the authors' knowledge, the
weak-localization magnetoresistances in ITO nanowires has previously been
reported only in one study in which the electrical measurements were performed
on two-probe, rather than four-probe, individual nanowire devices
\cite{Chiquito07}. In that experiment, the authors reported a very short
electron dephasing (phase-breaking) length, $L_\varphi$, of 31 (15) nm at 10
(100) K.

In this work, we have investigated low-temperature magnetoresistances in the
weak-(anti)localization effect in one low-resistivity and one high-resistivity
ITO nanowires, by employing the four-probe configuration method. The electron
dephasing length $L_\varphi$ and the spin-orbit scattering length, $L_{\rm
so}$, are experimentally extracted. We found a very long $L_\varphi$(0.25 K)
$\approx$ 520 nm, which is considerably longer than most electron dephasing
lengths reported in previous studies of as-grown nanowires. Furthermore, we
demonstrate that the strength of spin-orbit coupling can be adjusted by
modifying the level of disorder or electron mean free path, $l$, in the
nanowire. As a consequence, both the weak-localization effect (which is
characterized by negative magnetoresistances) and the weak-antilocalization
effect (which is characterized by positive magnetoresistances) can be `tuned'
and visualized in ITO single nanowires in low magnetic fields.

\section{Experimental method}

ITO (In$_{2-x}$Sn$_x$O$_{3-y}$) nanowires were synthesized on Si substrates by
the standard thermal evaporation method, as described previously
\cite{Chiu-nano09ITO}. X-ray diffraction (XRD, MAC Science MXP-18) and
high-resolution transmission electron microscopy (TEM, JEOL JEM-2010FX)
studies indicated that the nanowires were single crystalline, possessing a
cubic bixbyite structure, and grew along the [100] direction. Energy
dispersive x-ray spectroscopy (EDS, JOEL JEM-2010FX) and inductively coupled
plasma atomic emission spectrometry (ICP-OES, Perkin Elmer Optima-3000DV) were
utilized to determine the chemical compositions of the nanowires. We obtained
an Sn/In weight ratio of $\sim$ 4.4\%, confirming that Sn was effectively
incorporated into the ITO nanowires \cite{Chiu-nano09ITO}.

Both the electron-beam lithography and focused-ion beam (FIB) techniques were
employed to make four-probe individual nanowire devices for magnetoresistance
measurements. We first made micron-sized Ti/Au (10/60 nm) metal pads and
coordinate marks by utilizing the standard photolithography and lift-off
technique on a SiO$_2$ layer (400 nm thick) capped Si substrates. ITO
nanowires were then dispersed on the SiO$_2$/Si substrates. Individual
nanowires with diameters of $\sim$ 60--70 nm were identified and electrically
connected to the pre-patterned micron-sized Ti/Au metal pads by using the
standard electron-beam lithography and lift-off technique (for the ITO-r
nanowire device). In order to make good ohmic contacts to the nanowire, O$_2$
plasma was employed to clean the substrate before the Cr/Au (10/100 nm)
electrodes contacting the nanowire were deposited by the thermal evaporation
method. A scanning electron microscope (SEM) image of the ITO-r nanowire is
shown in the inset of Fig. \ref{fig2}.

To fabricate the four-probe ITO-f individual nanowire device, we employed the
FIB technique. In an FEI Dual-Beam NOVA 200 FIB instrument, we applied the
Ga$^+$-beam-induced deposition method with a methyl cyclopentadienyl trimethyl
platinum (CH$_3$)$_3$Pt(C$_{\rm p}$CH$_3$) injector to selectively deposit Pt
metal for connecting the ITO nanowire with the Ti/Au pads. The Ga$^+$-ions
were accelerated to 30 kV at 10 pA and injected into the nanowire during the
Pt deposition. The dimensions of the Pt contacting electrodes were set 80 nm
in width and 100 nm in thickness, but the actual width was about 300 nm and
the actual thickness was not determined. An SEM image of the four-probe ITO-f
nanowire device is shown in the inset of Fig. \ref{fig4}.

The individual nanowire devices were placed on a sample holder which was
situated inside a dark vacuum can. The vacuum can was mounted on an Oxford
Heliox $^3$He cryostat for magnetoresistance measurements from 40 K down to
0.25 K. The cryostat was equipped with a 4-T superconducting solenoid, and the
temperature was monitored with calibrated RuO$_2$ and Cernox thermometers. An
AC resistance bridge (Linear Research LR-700) was employed for the measurement
of magnetic field dependent resistance, $R(B)$, at various fixed temperatures.
In all cases, the magnetic fields were applied perpendicular to the nanowire
axis. A small measuring current ($\lesssim$ 30 nA) was employed to avoid Joule
heating. Linearity in the current-voltage curves was repeated checked and
ensured.

\section{Results and discussion}

\begin{figure}
\includegraphics[angle=270,scale=0.32]{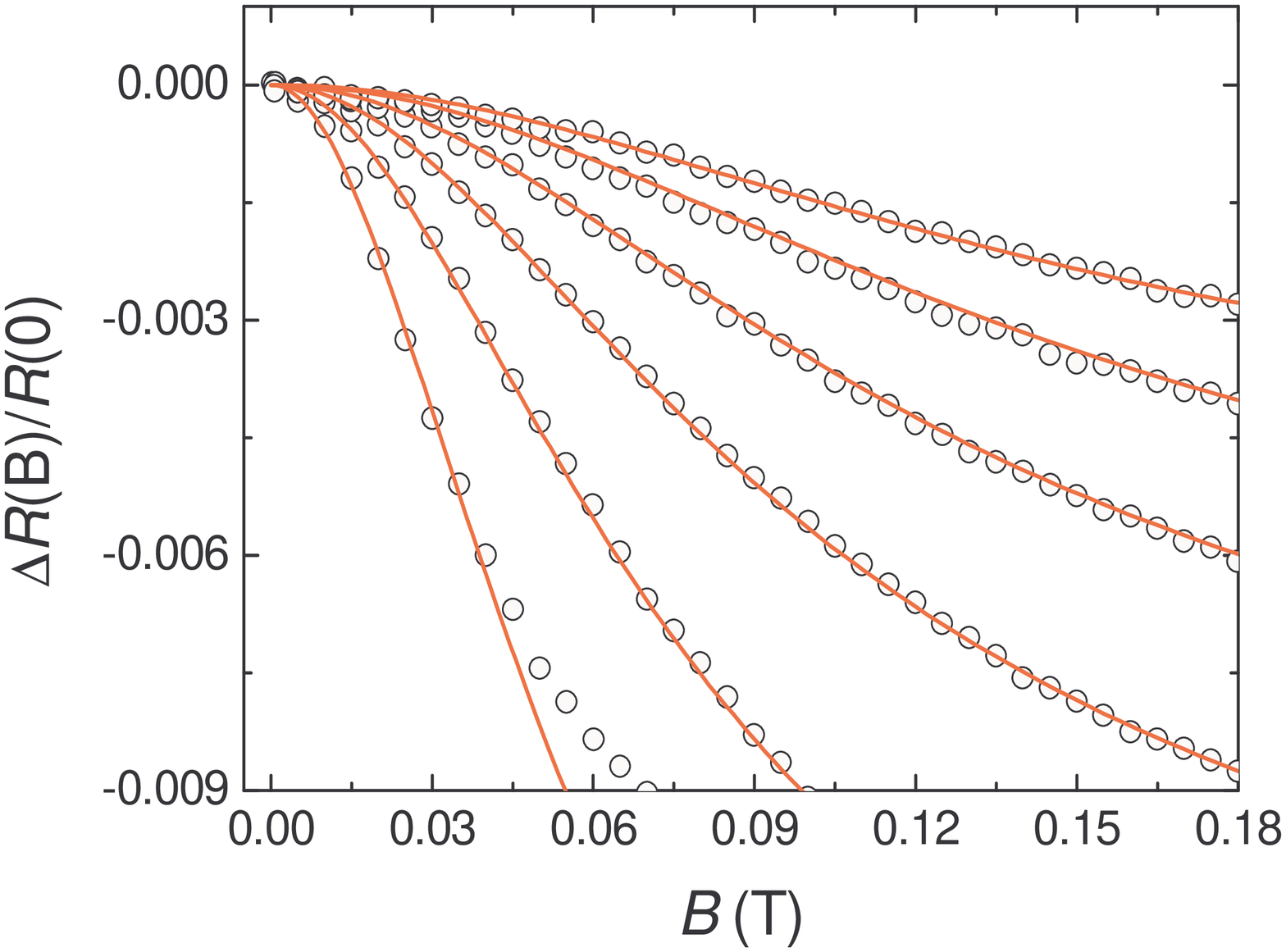}
\caption{\label{fig1} Normalized magnetoresistance as a function of
perpendicular magnetic field of the ITO-r nanowire at (from bottom up): 0.25,
5.0, 12, 20, 30, and 40 K. The symbols are the experimental data and the solid
curves are the theoretical predictions of Eq. (\ref{eq1}).}
\end{figure}

Figure~\ref{fig1} shows the normalized magnetoresistance, $\triangle R(B)
/R(0) = [R(B) - R(0)]/R(0)$, as a function of magnetic field at several
temperatures of the low-resistivity ITO-r nanowire. This figure reveals that
the magnetoresistances are always negative, implying a very weak spin-orbit
scattering rate relative to the inelastic electron scattering rate even down
to 0.25 K in this particular nanowire. Our measured low-field
magnetoresistances were least-squares fitted to the one-dimensional
weak-localization theoretical prediction \cite{Altshuler87,Birge03}
\begin{eqnarray}
\frac{\triangle R(B)}{R(0)} & = & \frac{e^2}{\pi \hbar} \frac{R}{L} \Biggl\{
\frac{3}{2} \Biggl[ \Biggl( \frac{1}{L_\varphi^2} + \frac{4}{3L_{so}^2} +
\frac{W^2}{3 L_B^4} \Biggr)^{-1/2} \nonumber \\ & - & \Biggl( \frac{1}{L_\varphi^2} +
\frac{4}{3L_{so}^2}
\Biggr)^{-1/2} \Biggr] \nonumber \\
& - & \frac{1}{2} \Biggl[ \Biggl( \frac{1}{L_\varphi^2} + \frac{W^2}{3 L_B^4}
\Biggr)^{-1/2} - L_\varphi \Biggr] \Biggr\} \,, \label{eq1}
\end{eqnarray}
where $R$ is the resistance of a nanowire of width $W$ and length $L$, $L_B =
\sqrt{\hbar /eB}$ is the magnetic length ($\hbar$ is the Planck constant
divided by $2\pi$, and $e$ is the electronic charge), $L_\varphi = \sqrt{D
\tau_\varphi}$ is the electron dephasing length, $L_{\rm so} = \sqrt{D
\tau_{\rm so}}$ is the spin-orbit scattering length, and $D = v_F^2 \tau_e/3$
is the electron diffusion constant \cite{diffusion} ($v_F$ being the Fermi
velocity \cite{Fermi}, and $\tau_e$ being the electron mean free time). Here
$\tau_\varphi$ and $\tau_{\rm so}$ are the electron dephasing time and the
spin-orbit scattering time, respectively. The spin-orbit scattering length
(time) is a temperature independent quantity whose size, relative to the
inelastic scattering strength, determines the sign of the
weak-(anti)localization effects in the low-field magnetoresistance
\cite{Altshuler87,Bergmann82}.

In Fig. \ref{fig1}, the symbols are the experimental data and the solid curves
are the theoretical predictions of Eq. (\ref{eq1}). This figure clearly
indicates that the measured magnetoresistances can be well described by the
one-dimensional weak-localization theory in the wide temperature interval of
0.25--40 K. Therefore, the important characteristic electron scattering
lengths, $L_\varphi$ and $L_{\rm so}$, can be accurately extracted. In this
low-resistivity nanowire, we obtained an extraordinarily long $L_{\rm so}$
($\tau_{\rm so}$). In other words, the spin-orbit interaction is negligibly
weak (absent), to within our experimental uncertainty, in this sample. [We
estimate the length $L_{\rm so} \gtrsim$ 0.5 $\mu$m, or the time $\tau_{\rm
so} \gtrsim 1 \times 10^{-10}$ s, in this nanowire (see below). It is worth
noting that this spin-orbit scattering length is comparable to or even longer
than that in relatively clean Al films and wires \cite{Santhanam87}.]

\begin{figure}
\includegraphics[angle=270,scale=0.34]{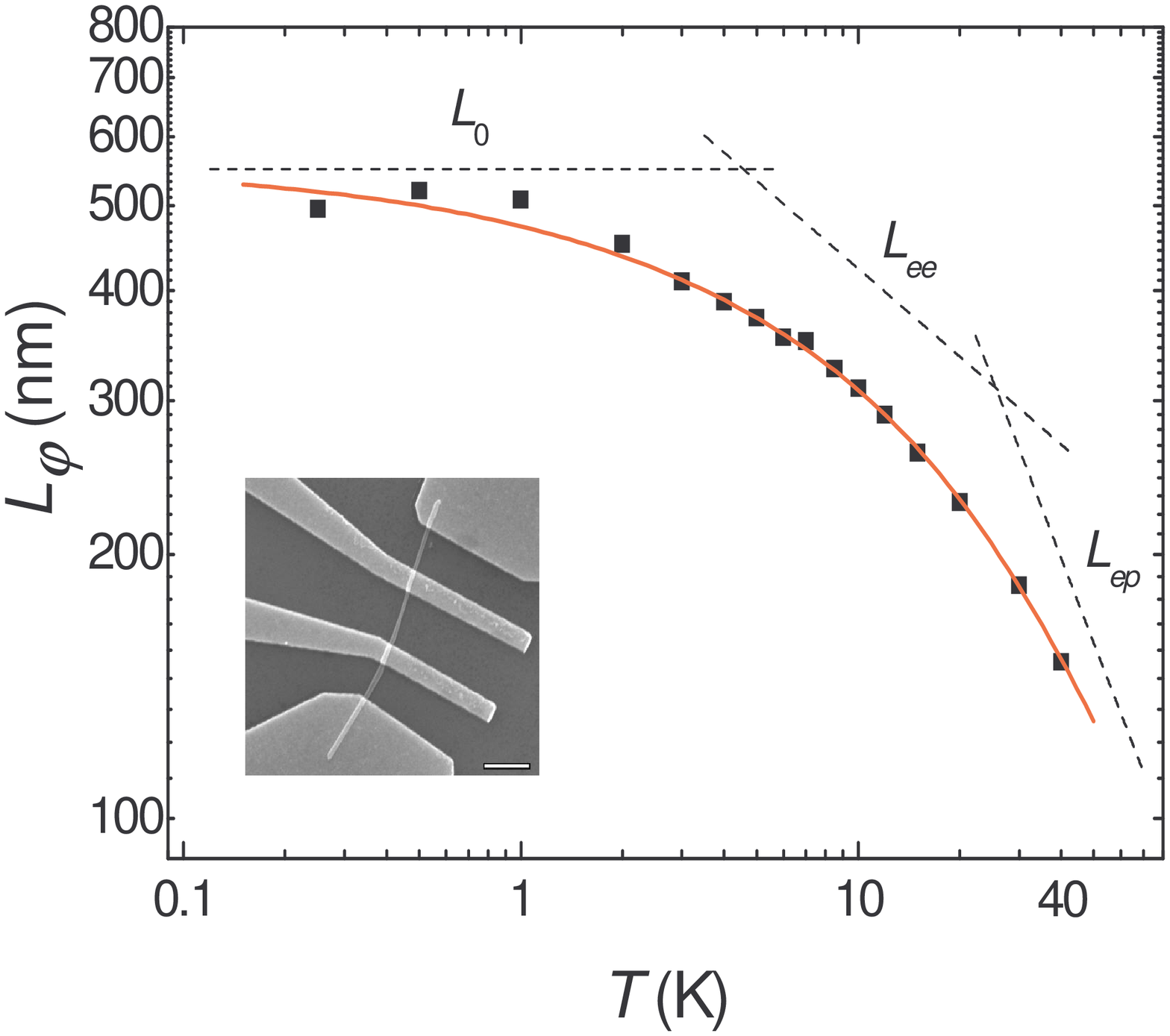}
\caption{\label{fig2} Electron dephasing length as a function of temperature
for the ITO-r nanowire. The symbols are the experimental data and the solid
curve is the theoretical prediction of Eq. (\ref{eq2}). The three straight
dashed lines represent our fitted values of $L_0$, $L_{ee}$ and $L_{ep}$, as
indicated. The inset shows an SEM image of this individual nanowire device.
The scale bar is 1 $\mu$m.}
\end{figure}

Figure~\ref{fig2} shows our extracted $L_\varphi$ as a function of temperature
for the low-resistivity ITO-r nanowire. The inset shows an SEM image of this
individual nanowire device. This figure indicates that $L_\varphi$ increases
from 150 nm to 520 nm as the temperature decreases from 40 K down to 0.25 K
\cite{wire-length}. This result bears significant implications. As a
consequence of such a long $L_\varphi$, the weak-localization behavior is
purely one-dimensional in this particular nanowire device over our whole
measurement temperature range. In fact, it is clear that the one-dimensional
weak-localization behavior will persist up to much above 40 K. In contrast, in
previous studies of individual nanowires made of ITO
\cite{Chiquito07,Berengue09}, and other materials such as InAs
\cite{Liang09,Liang10} and P-doped Si \cite{Rueb-prb07}, the magnetoresistance
measurements were performed on two-probe configurations and the obtained
$L_\varphi$ were often short ($\lesssim$ 100 nm at 1 K). Thus, a crossover of
the device dimensionality from one to three, in regard to the
weak-localization effect, inevitably happens already at low temperatures of a
few degrees of Kelvin. Under such circumstances, the extracted values of
$L_\varphi$ ($\tau_\varphi$) above a few degrees of Kelvin are obviously
questionable.

\begin{table*}
\caption{\label{t1} Relevant parameters for two ITO individual nanowire
devices. $d$ is the nanowire diameter, $L$ is the nanowire length between the
two voltage electrodes in the four-probe configuration, $\ell$ is the electron
mean free path, $D$ is the electron diffusion constant, $\tau_{\rm so}$ is the
spin-orbit scattering time, $\tau_0 = \tau_\varphi (T \rightarrow 0$ K) is the
electron dephasing time as $T$ approaches zero, and $A_{ep}$ is the
electron-phonon scattering strength. The values of $\ell$ and $D$ were
calculated for 10 K. Since a dimensionality crossover occurs at $\approx$12 K
in the ITO-f nanowire, the value of $A_{ep}$ for this device is an approximate
upper bound.}

\begin{ruledtabular}
\begin{tabular}{lcccccccccc}

%\br

Sample & $d$ & $L$ & $\rho$(300\,K) & $\rho$(10 K) & $\ell$ & $D$ & $\tau_e$ &
$\tau_{\rm
so}$ & $\tau_0$ & $A_{ep}$ \\

 & (nm) & ($\mu$m) & ($\mu \Omega$ cm) & ($\mu \Omega$ cm) & (nm) & (cm$^2$/s)
& (fs) & (ps) & (ps) & (K$^{-2}$ s$^{-1}$) \\ \hline

%\mr

ITO-r & 60 & 1.5 & 202 & 185 & 9.3 & 25 & 12 & $\gtrsim$100 & 120 & $4.0 \times 10^7$ \\
ITO-f & 72 & 2.8 & 997 & 1030 & 2.9 & 5.9 & 4.2 & 26 & 69 & $5 \times 10^8$ \\

%\br
\end{tabular}
\end{ruledtabular}
\end{table*}

In a weakly disordered conductor, the electron dephasing time is given by
\cite{Lin-jpcm02}
\begin{equation}
\frac{1}{\tau_\varphi (T)} = \frac{1}{\tau_0} + \frac{1}{\tau_{ee}(T)} +
\frac{1}{\tau_{ep}(T)} \,, \label{eq2}
\end{equation}
where $\tau_0$  is a constant, whose origins (paramagnetic impurity
scattering, dynamical structural defects, etc.) are a subject of elaborate
theoretical and experimental investigations
\cite{Birge03,Lin-prb87b,Mohanty97,Lin-JPSJ03,Huang-prl07}. The
electron-electron relaxation rate, $\tau_{ee}^{-1}$, in low-dimensional
disordered conductors is known to dominate $\tau_\varphi^{-1}$ in an
appreciably wide temperature interval \cite{Altshuler87,Lin-jpcm02}. The
electron-phonon scattering rate, $\tau_{ep}^{-1}$, will become important at
high measurement temperatures. We use the standard expression: $\tau_{ep}^{-1}
= A_{ep}T^2$ in the quasi-ballistic limit of $q_Tl > 1$ (which is pertinent to
the present study) \cite{Sergeev-prb00,Zhong-prl10}, where $q_T$ is the
wavenumber of a thermal phonon and $l$ is the electron mean free path. The
electron-electron relaxation rate was fitted in the form $\tau_{ee}^{-1} =
A_{ee} T^p$. As it is shown in Fig. \ref{fig2}, the dependence $L_\varphi =
\sqrt{D \tau_\varphi}$ is well fitted by Eq. (\ref{eq2}) with four adjustable
parameters $\tau_0$, $A_{ee}$, $p$, and $A_{ep}$. Our fitted values of
$\tau_0$ and $A_{ep}$ are listed in Table \ref{t1}. It should be stressed that
we obtain an exponent of temperature $p \approx 0.68 \pm 0.10$, which is very
close to the theoretical value (= 2/3) of the Nyquist electron-electron
scattering rate in one dimension. Moreover, our fitted value of $A_{ee}$ is in
good agreement with the theoretical prediction of
\cite{Altshuler87,Lin-jpcm02,Birge03}
\begin{equation}
\frac{1}{\tau_{ee}} = \Biggl( \frac{e^2 \sqrt{D} R k_BT}{2 \sqrt{2} \hbar^2 L}
\Biggr)^{2/3} = A_{ee}T^{2/3} \,. \label{eq3}
\end{equation}
Equation (\ref{eq3}) predicts a scattering rate with the coefficient $A_{ee}
\approx 5.1 \times 10^9$ K$^{-2/3}$ s$^{-1}$ in the ITO-r nanowire, while
experimentally we obtain $A_{ee} \approx 3.0 \times 10^9$ K$^{-2/3}$ s$^{-1}$.
Also plotted in Fig. \ref{fig2} are our fitted values of the three dephasing
contributions $L_0 = \sqrt{D \tau_0}$, $L_{ee} = \sqrt{D \tau_{ee}}$, and
$L_{ep} = \sqrt{D \tau_{ep}}$, as indicated. This figure indicates that the
electron-phonon scattering becomes comparable to the electron-electron
scattering at $\approx$ 25 K.

\begin{figure}
\includegraphics[angle=270,scale=0.32]{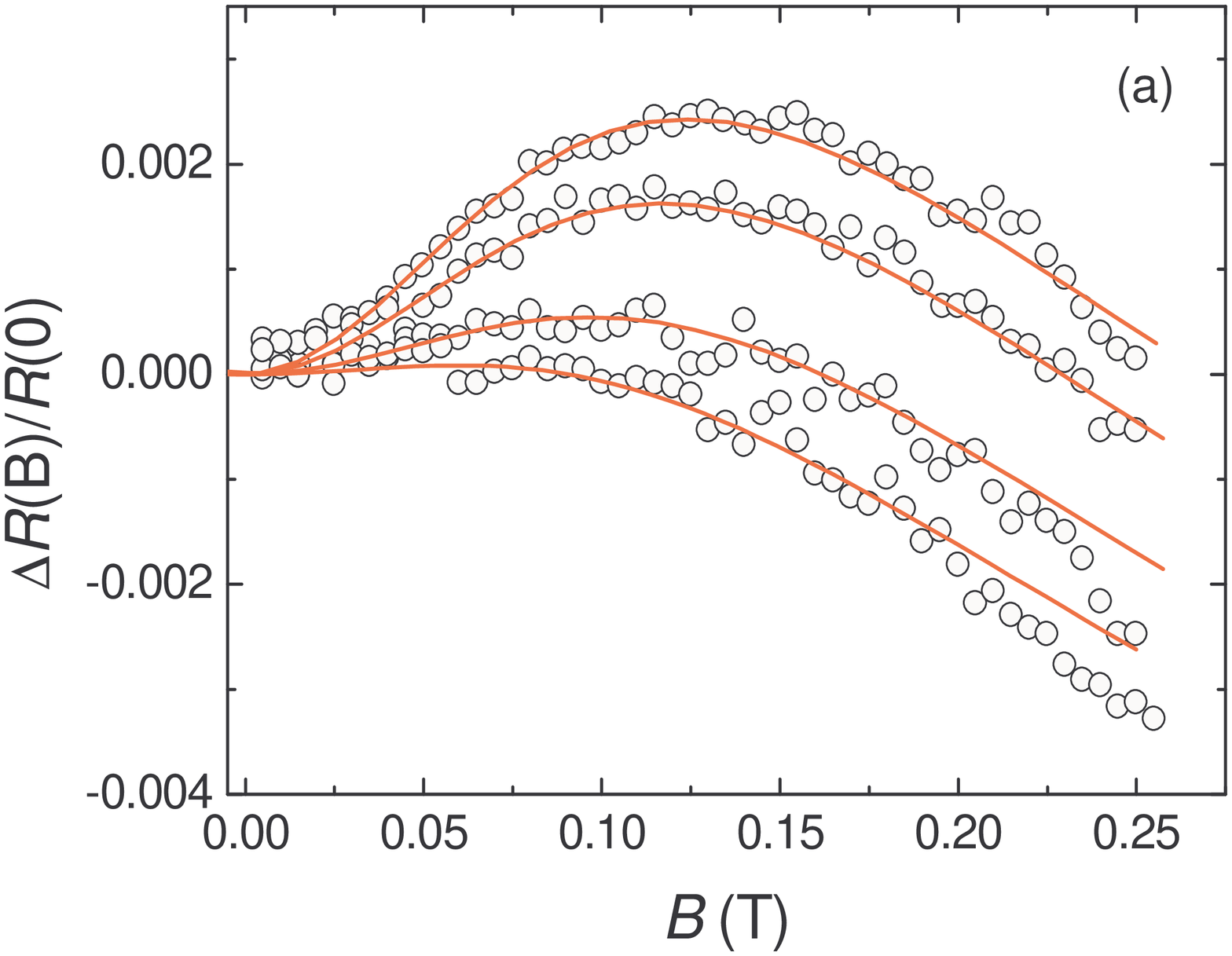}
\includegraphics[angle=270,scale=0.32]{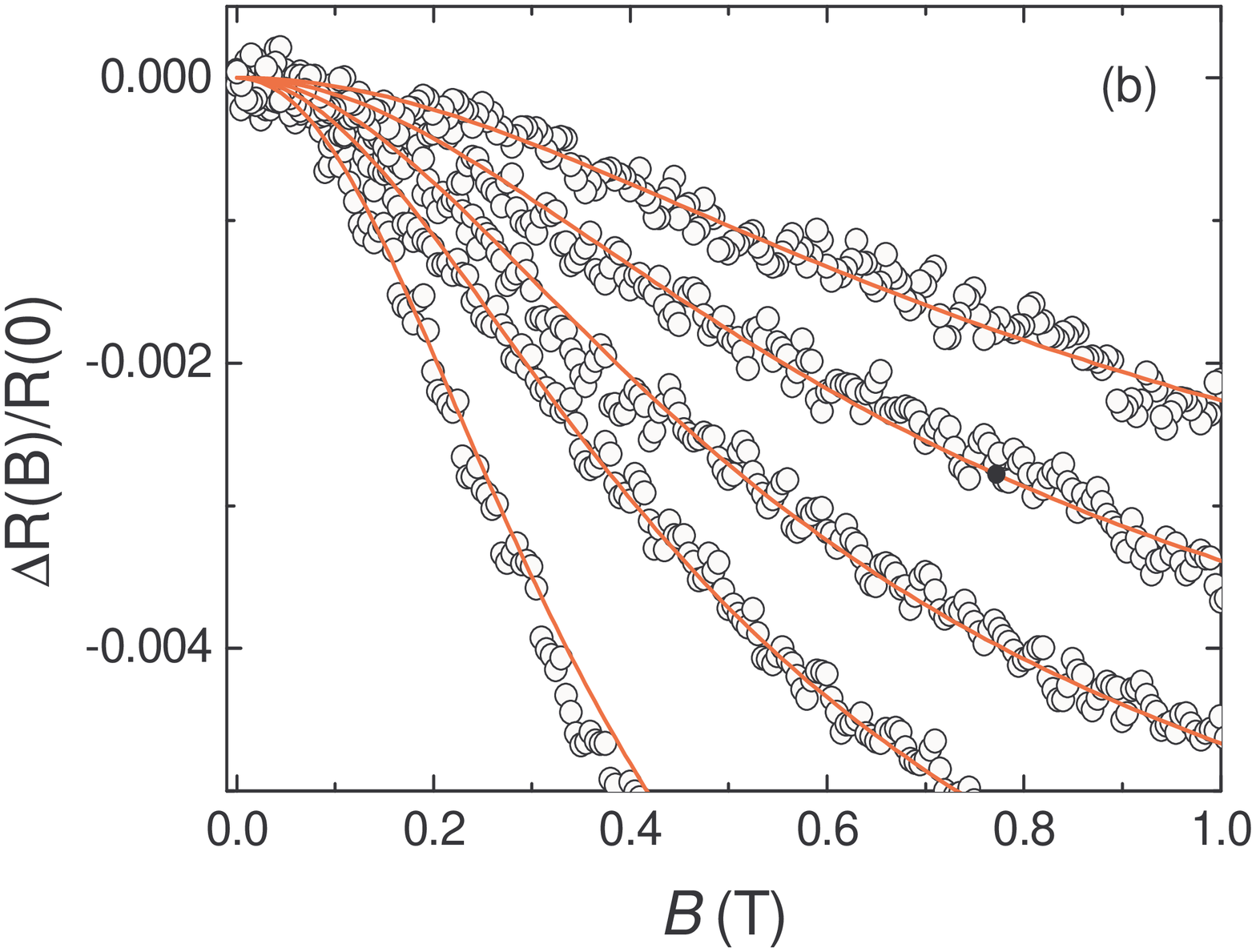}
\caption{\label{fig3} (a) Normalized magnetoresistance as a function of
perpendicular magnetic field of the ITO-f nanowire at (form top down) 0.26,
1.0, 2.0, and 4.0 K. (b) Normalized magnetoresistances as a function of
perpendicular magnetic field at (from bottom up) 8.5, 18, 22, 30, and 40 K.
The symbols are the experimental data and the solid curves are the theoretical
predictions of Eq. (\ref{eq1}).}
\end{figure}

Figure \ref{fig3}(a) shows the normalized magnetoresistances of the
high-resistivity ITO-f nanowire \cite{resistivity} as a function of magnetic
field at four temperatures between 0.26 and 4 K. Figure \ref{fig3}(b) shows the
normalized magnetoresistances as a function of magnetic field at five
temperatures between 8.5 and 40 K. These figures reveal that the
magnetoresistances are negative in all fields at temperatures above $\sim$ 6
K. Nevertheless, below 4 K, the magnetoresistances are positive in low
magnetic fields, changing to negative magnetoresistances in higher magnetic
fields. The positive contribution is more pronounced at lower temperatures.
This marked behavior can be readily understood in terms of the
quantum-interference effect in the presence of a notable spin-orbit scattering
rate relative to the inelastic electron-electron and electron-phonon
scattering rates. Under such conditions, positive magnetoresistances arise
owing to the weak-antilocalization effect \cite{Altshuler87,Lin-jpcm02}.

The presence of a notable spin-orbit scattering rate in the ITO-f nanowire can
be explained as follows. In this nanowire, the resistivity [$\rho$(10\,K) =
1030 $\mu \Omega$ cm)] is considerably higher than that (185 $\mu \Omega$ cm)
in the ITO-r nanowire. Therefore, the spin-orbit interaction in this nanowire
should be significantly enhanced according to the (approximate) relation
\cite{Santhanam87,Abrikosov,Bergmann85,Bergmann10}: $1/\tau_{\rm so} \propto
Z^4/\tau_e$, where $Z$ is the atomic number of the relevant scatterer. A
comparatively high resistivity in the ITO-f nanowire implies a relatively
short $\tau_e$, which should lead to a large $\tau_{\rm so}^{-1}$.
Quantitatively, we obtain $L_{\rm so} \approx$ 125 nm in this nanowire (see
below). The present experiment suggests that one can tailor both the
weak-localization and weak-antilocalization effects in ITO nanowires by
carefully controlling the level of atomic imperfections (point defects and/or
heavy impurities). This observation provides valuable information about the
feasible implementation of future quantum-interference nanodevices. In
addition to a long dephasing length (leading to sensitive quantum transport
effects), the capability of tuning spin-orbit coupling could be advantageous
for the future realization of spintronic devices \cite{Zutic04}.

\begin{figure}
\includegraphics[angle=270,scale=0.34]{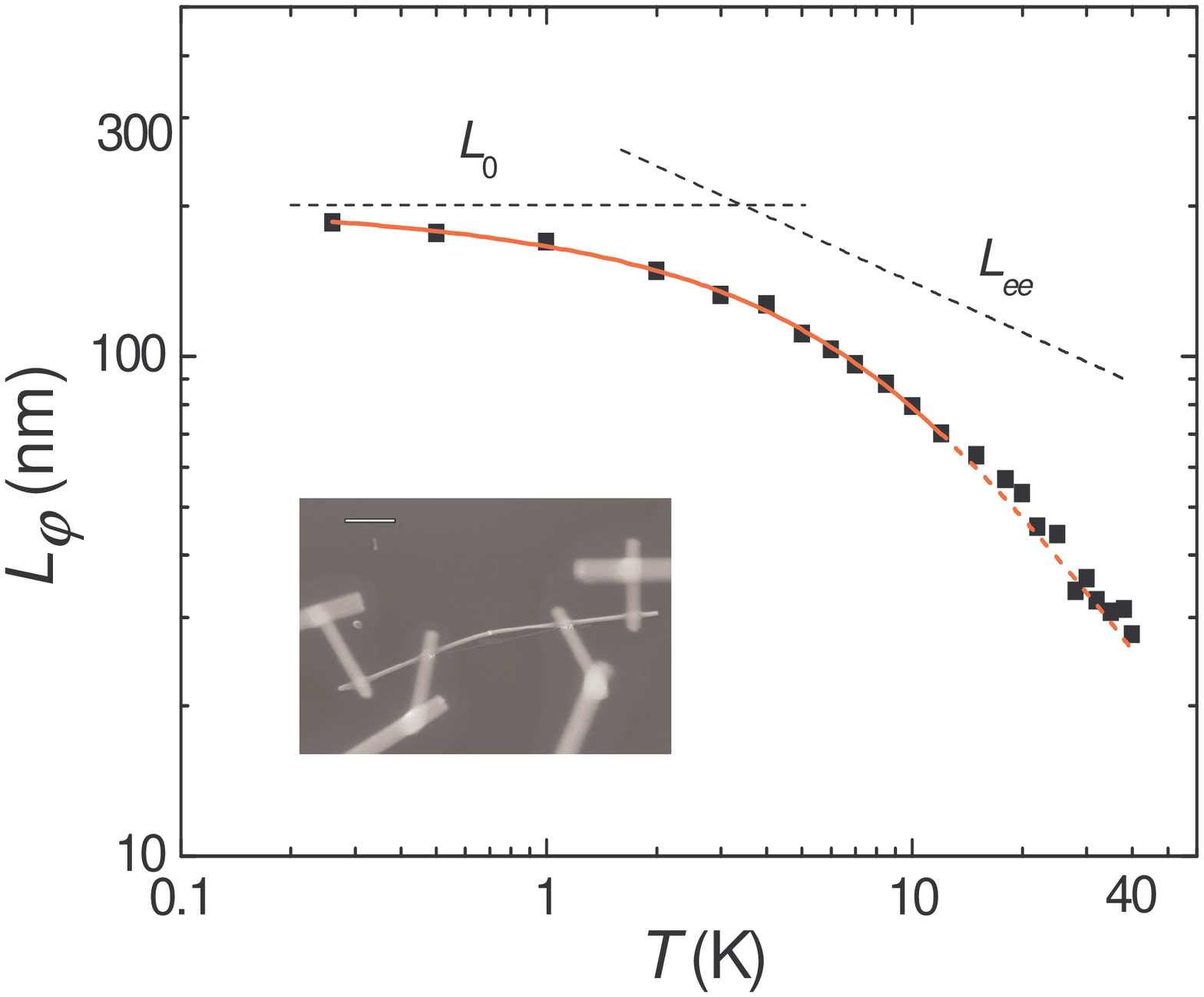}
\caption{\label{fig4} Electron dephasing length as a function of temperature
for the ITO-f nanowire. The symbols are the experimental data and the solid
curve is the theoretical prediction of Eq. (\ref{eq2}). The extrapolated
dashed curve above 12 K is a guide to the eye. The two straight dashed lines
represent our fitted values of $L_0$ and $L_{ee}$, as indicated. The inset
shows an SEM image of this individual nanowire device. The scale bar is 1
$\mu$m.}
\end{figure}

Our measured magnetoresistances in the ITO-f nanowire can be well described by
the predictions of Eq. (\ref{eq1}) [the solid curves in Figs. \ref{fig3}(a)
and \ref{fig3}(b)], and the extracted $L_\varphi$ as a function of temperature
is plotted in Fig. \ref{fig4}. The inset shows an SEM image of this individual
nanowire device. Inspection of Fig. \ref{fig4} indicates that $L_\varphi$ in
this nanowire is considerably shorter than that in the ITO-r nanowire. This
can be explained by the fact that $L_\varphi$ is suppressed by a high level of
disorder, i.e., a short mean free path, in the ITO-f nanowire. [Note that Eq.
(3) predicts that $1/\tau_{ee} \propto 1/\sqrt{\tau_e}$, and $1/\tau_{ep}
\propto 1/\tau_e$ in the quasi-ballistic limit
\cite{Sergeev-prb00,Zhong-prl10}.] Our result was fitted to Eq. (\ref{eq2})
(the solid curve in Fig. 4) and the adjustable values of $\tau_0$ and $A_{ep}$
are listed in Table \ref{t1}. Again, we obtain the exponent of temperature $p
\approx 0.69 \pm 0.12$ in $\tau_{ee}^{-1}$, being very close to the
theoretical value of 2/3. In addition, our fitted value of the one-dimensional
electron-electron scattering coefficient $A_{ee} = 6.4 \times 10^9$ K$^{-2/3}$
s$^{-1}$ is in excellent agreement with the theoretical prediction of $7.9
\times 10^9$ K$^{-2/3}$ s$^{-1}$ from Eq. (\ref{eq3}). Such close agreement
between experiment and theory most likely arises from the intricate material
fact that, though being a doped oxide, ITO possesses a free-carrier-like
(i.e., parabolic) electronic band structure
\cite{Chiu-nano09ITO,Li-JAP04,Mryasov01}. We notice in passing that, in
another high-resistivity nanowire with $d$ = 110 nm and $\rho$(10 K) = 1690
$\mu \Omega$ cm, we obtained $L_\varphi$(0.25 K) = 170 nm and $L_{\rm so}$ =
95 nm. These values are in good consistency with the corresponding values in
the ITO-f nanowire.

It should be noted that in the ITO-f nanowire, our extracted value of
$L_\varphi$ becomes smaller than the nanowire diameter at a temperature of
$\sim$ 12 K. That is, a dimensionality crossover of the weak-localization
effect nominally begins to set in at $\sim$ 12 K in this particular nanowire
device. Strictly speaking, the measured magnetoresistances should then be
least-squares fitted with the three-dimensional weak-localization theoretical
predictions \cite{Lin-jpcm02,Wu-prb94} in order to extract $L_\varphi$.
However, we found that between 12 and 40 K, our measured magnetoresistances
can still be satisfactorily described by Eq. (\ref{eq1}) [see Fig.
\ref{fig3}(b)], while the three-dimensional weak-localization theory does not
apply \cite{3D}. In any case, the extracted value of $L_\varphi$ above 12 K in
this nanowire device should be considered only qualitative and treated with
caution. (Hence, in Fig. \ref{fig4}, we only plotted our fitted values of the
two contributions $L_0$ and $L_{ee}$.)

\section{Conclusion}

We have quantitatively measured the electron dephasing length in two
individual ITO nanowires between 0.25 and 40 K. We observe that the electron
dephasing length is very long, reaching 520 nm at 0.25 K in a low-resistivity
ITO nanowire. As a consequence, the sample demonstrates strict one-dimensional
weak-localization effect up to temperatures above 40 K. In a high-resistivity
nanowire, the spin-orbit coupling is enhanced due to a short electron mean
free path, manifesting the weak-antilocalization effect at temperatures below
$\sim$ 4 K. These observations provide strong indications that robust
quantum-interference effects can be realized, and tunable, in ITO nanowires by
controlling differing levels of atomic defects and impurities.

\begin{acknowledgments}

The authors are grateful to C. Y. Wu for careful reading of the manuscript,
and F. R. Chen and J. J. Kai for providing us with the ITO nanowires used in
this study. This work was supported by Taiwan National Science Council through
Grant No. NSC 98-2120-M-009-004, and by the MOE ATU Program.

\end{acknowledgments}


\begin{thebibliography}{99}
\bibitem{diffusion} While our ITO nanowires are one-dimensional with regard
to the quantum-interference weak-(anti)localization effect, they are
three-dimensional with regard to the classical Boltzmann transport, i.e., the
electron mean free path $l < d$, the diameter of nanowire.
\bibitem{Fermi} Our Fermi velocities were evaluated by using valus of $E_F$
= 0.7 and 0.4 eV in the ITO-r and ITO-f nanowires, respectively. See
Refs. \onlinecite{Chiu-nano09ITO} and \onlinecite{Li-JAP04}.
\bibitem{resistivity} Recent experiment (Ref. \onlinecite{Wan06}) reported that
the resistivities of ITO nanowires grown under nominally similar conditions
could vary widely, from $\sim$ 100 $\mu \Omega$ cm to values higher than
several thousands of $\mu \Omega$ cm. In our case, the resistivity differs by
a much smaller factor of $< 10$ (Ref. \onlinecite{Chiu-nano09ITO}).
\bibitem{3D} Least-squares fits of our measured magnetoresistances with a
combined one-dimensional and three-dimensional weak-localization expression
did not lead to acceptable values of the adjustable parameters. A
weak-localization theory in the crossover regime is yet to be established.
\end{thebibliography}
\end{document}